\documentclass[11pt,twoside]{article}
\usepackage{asp2004}
\usepackage{psfig}
\usepackage{epsf}
\usepackage{graphics}
\usepackage{graphicx}
\usepackage{lscape}
\begin{document}

\title{ New release of the ELODIE library}
\author{Ph. Prugniel (1) \& C. Soubiran(2)}
\affil{(1) CRAL-Observatoire de Lyon and  GEPI-Observatoire de Paris\\
(2) L3AB-Observatoire de Bordeaux}

\begin{abstract}
 We present ELODIE.3, an updated release of the ELODIE
 library  originally published in Prugniel \& Soubiran
 (2001). It is part of the
 spectrophotometric resources  available in the HyperLeda database where the
 spectra can be visualized or  downloaded in FITS format. The ELODIE
 library includes 1962 spectra of  1388 stars obtained with the ELODIE
 spectrograph at the Observatoire de  Haute-Provence 193cm telescope in
 the wavelength range 400 to 680 nm. It  provides a large coverage of
 atmospheric parameters : Teff from 3000 K  to 60000 K, log g from -0.3
 to 5.9 and [Fe/H] from -3.2 to +1.4. The  library is given at two
 resolutions: R=42000, with the flux normalized  to the
 pseudo-continuum, R=10000 calibrated in physical flux (reduced  above
 earth atmosphere) with a broad-band photometric precision of 2.5\%  and
 narrow-band precision of 0.5\%. A grid of interpolated spectra is  also
 given, which has been used for synthesis of stellar population with
 PEGASE-HR (Le Borgne et al. 2004). Compared to
 the previous version we have doubled the size of the library, corrected 
 some identification errors, improved the external stellar parameters and
 the reduction procedure. A detailed 
 description is available at the ELODIE.3 website :\\ 
 
 \noindent
http://www.obs.u-bordeaux1.fr/m2a/soubiran/elodie\_library.html

\end{abstract}
\thispagestyle{plain}

\section{ Content of the library}
Fig. 1 and 2 show the distribution of the 1388 stars of the library in
the HR diagram and Teff - [Fe/H] plane. Stellar parameters and useful 
data have been 
searched in the literature or determined for all stars. They are described 
and can be retreived from the ELODIE.3 website.

\begin{figure*}[ht]
\includegraphics[width=11cm]{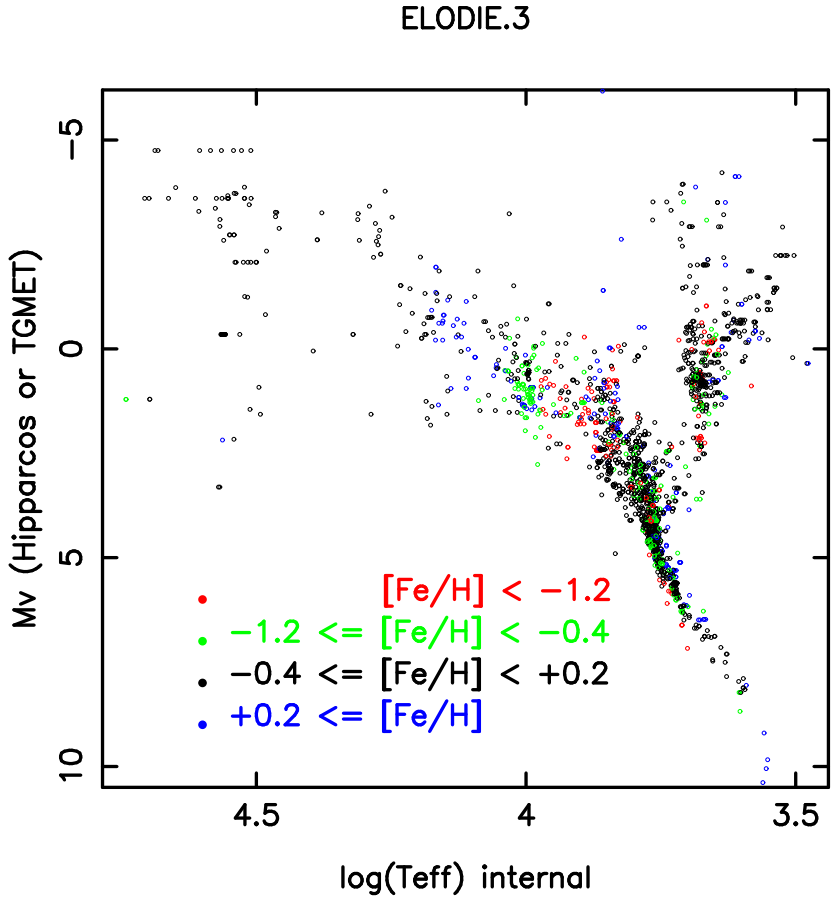}
\caption{Distribution of the ELODIE.3 library
in the HR diagram, with three metallicity bins differentiated with symbols 
of different colors
}
\label{TeffMv}
\end{figure*}

\begin{figure*}[ht]
\includegraphics[width=11cm]{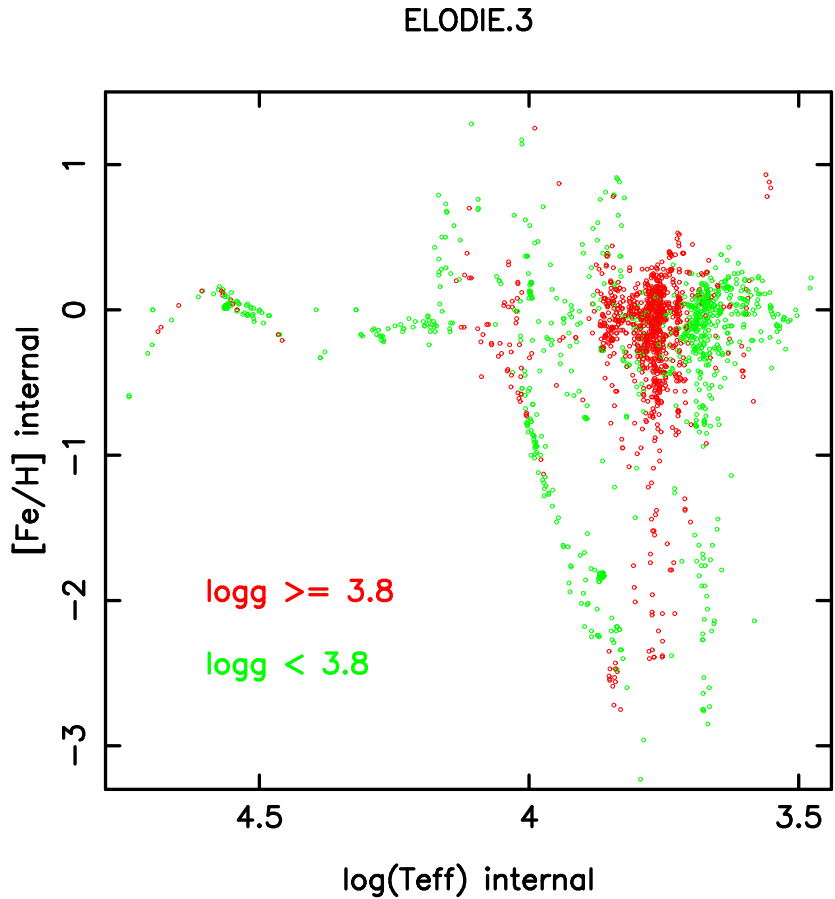}
\caption{Distribution in the Teff - [Fe/H] 
plane for dwarfs and giants
}
\label{TeffFeH}
\end{figure*}

\section{ Reduction procedure}
Since the first version of the library, the reduction procedure described 
in Prugniel \& Soubiran (2001) has been improved in several aspects but the 
general 
philosophy remains the same. The basic steps in the reduction are to

\begin{itemize}
\item correct the different orders in the spectra for the blaze effect and 
connect the orders together

\item mask the telluric lines and spikes due to cosmic rays : in the new 
version, the mask of the spikes due to cosmic rays has been suppressed 
because it was found to alter real features. Because of it, in the previous 
version of the library Lick indices measured on our spectra presented a slope 
compared to measurements on Jones spectra, the strongest features were found 
weaker than real (thanks to G. Worthey who stressed our attention on this 
important problem).

\item make the flux calibration

\item build an interpolator which in turn is used to generate the grid that 
feeds the population synthesis program (PegaseHR).
\end{itemize}
The strengthening of the spectra have been improved: artifacts near Balmer's 
lines were corrected. The wavelength interval has been extended by adding 5 
orders in the blue. It was 410-680 nm, it is now 400-680. The correction for 
telluric lines has been improved. We corrected some small bugs in the FITS 
keywords (in particular an unfortunate bad rounding of CRVAL1). \\
A full description of the reduction procedure can be found on the ELODIE.3 
website.

\section{  Data access}
The spectra are available as fits file. Three gzipped tarfiles, corresponding to 
the R=10000, R=42000 libraries and the grid can be downloaded, as well as 
the table of measurements
containing the stellar parameters estimated internally and the Lick indices. The
description of the files, calibration stes, specific keywords are described
on the ELODIE.3 website.

\end{document}